# Ionization induced plasma grating and its applications in strong-field ionization measurements

**Chaojie Zhang, Zan Nie, Yipeng Wu, Mitchell Sinclair, Chen-Kang Huang, Ken A Marsh and Chan Joshi**

Department of Electrical and Computer Engineering, University of California Los Angeles, Los Angeles, CA 90095, USA

E-mail : chaojiez@ucla.edu

**Abstract.** An ionization induced plasma grating can be formed by spatially selective ionization of gases by the interference of two intersecting ultra-short laser pulses. The density modulation of a plasma grating can approach unity since the plasma is produced only where the two pulses constructively interfere and ionization does not occur in destructive interference regions. Such a large density modulation leads to efficient Thomson scattering of a second ultra-short probe pulse once the Bragg condition is satisfied. By measuring the scattering efficiency, it is possible to determine the absolute electron density in the plasma grating and thereby deduce the ionization degree for a given neutral gas density. In this paper, we demonstrate the usefulness of this concept by showing two applications: ionization degree measurement of strong-field ionization of atoms and molecules and characterization of extremely low-density gas jets. The former application is of particular interest for ionization physics studies in dense gases where the collision of the ionized electron with neighboring neutrals may become important-sometimes referred to as many-body ionization; and the latter is useful for plasma-based acceleration that requires extremely low-density plasmas.

## Introduction

As the fourth state of matter, plasma or matter in ionized state, is known to have unique properties that differentiate it from gases, liquids or solids. For instance, since it is already ionized and therefore has a free charge density of electrons, plasma can support extremely large electric fields without the electrical breakdown limit faced by conventional accelerators, making the concept of plasma-based acceleration very attractive in terms of reducing the size and cost of future accelerators and light sources [1–3]. For similar reasons, plasmas have also been considered as novel optical elements (such as mirror [4,5], polarizer [6–8], waveplate [7,8], q-plate [9] and waveguide [10,11]) suitable for emerging ultrahigh intensity laser systems [12–17].

Due to the long-range interactions between charged particles, plasma intrinsically supports an enormous number of modes (waves), either self-generated or externally excited. Such modes have been shown to trap and accelerate electrons, positrons or ions [3,18,19]. They also provide new possibilities for generating radiation over a broad range of frequencies [20–24] and topological structures [25,26].

Interference of two or more lasers in plasmas can generate standing [27–29] or travelling electron plasma waves or ion waves [30] that can be thought of as phase gratings. Such gratings have been used to determine the nonlinear refraction index of the plasma via a process known as degenerate four wave mixing [31], controlled study of the transient filamentation instability

[32] and to redistribute energy between multiple laser beams in an indirect drive laser fusion target by a process called cross-beam energy transfer [33,34].

Standing plasma gratings can be generated in different ways. Historically, two degenerate frequency laser beams were overlapped in plasma at various angles to generate standing wave electric fields that had a variable wavelength. At small angles, the wavelength of the beating pattern is much longer than the laser wavelength, whereas in the extreme case of counter-propagating beams it is one half of the laser wavelength. For small wavelength gratings the dominant mechanism is the pondermotive force whereas, for long wavelength gratings the thermal force can dominate over the ponderomotive force [31]. The former leads to ponderomotive self-focussing/filamentation, while the latter leads to thermal self-focussing/filamentation. Both require the laser pulses to be long enough so that the plasma ions have time to move.

In this paper we consider plasma gratings formed by the interference of two degenerate but ultrashort (<50 fs, FWHM) laser pulses in a neutral gas where ionization is important but ion motion and collisional heating are not. In this case the plasma grating is generated by the process of strong-field ionization. Even if the peak intensity of each laser pulse is below the threshold for strong-field (barrier-suppression) ionization, ionization can occur via tunnelling through the suppressed potential barrier or via simultaneous absorption of multiple photons in the vicinity of the constructive interference of the electric fields, leaving the regions where destructive interference occurs unaffected. The magnitude of this standing wave electron density modulation can approach unity within a few laser cycles. This unique feature enables efficient scattering of external probe pulses once the Bragg condition is satisfied [35]. We note that while in the early phase of the evolution of this electron grating, the more massive ions remain stationary, as electrons move once the laser pulse has left the region the ions are dragged by the charge separation field.

In this paper, we describe experimental studies of ionization induced plasma grating and its applications. We start with the generation of a plasma grating by two interfering lasers to periodically ionize atomic and molecular gases and subsequently follow its temporal evolution. By measuring the time-resolved Thomson scattering signal of an external probe pulse scattered off the plasma grating, we observe the decay of the grating caused by ambipolar expansion of the plasma. We then use the Thomson scattering method to determine the intensity dependence of the degree of ionization of atomic and molecular gases at several neutral gas pressures. We show that the measured ionization degree agrees well with the calculation using the Perelomov-Popov-Terent'ev (PPT) model [36]. Following that, we demonstrate a second application of the method in characterizing an extremely low-density supersonic gas jet that is needed for generating mm-scale wakes in plasmas driven by ps laser pulses.

**Experimental setup**

The experimental setup is shown in Fig. 1(a). A high rep-rate (up to 1 kHz) Ti:Sapphire laser system was used to deliver two pump laser pulses with central wavelength $\lambda_0 = 0.8$ μm that intersect each other at a small angle of either 12.8° or 16.5°, which gives intensity interference pattern with a spacing of 3.6 or 2.8 μm. Here the spacing is related to the wavelength of the pump beam and the intersecting angle $\theta$ through $\Lambda = \frac{\lambda_0}{2\sin(\theta/2)}$. The two pump lasers were focused using identical f=750 mm lenses to achieve a spot size of ~50 μm at the interaction point (IP). The pulse duration of the two pulses were measured to be ~43 fs at the IP.

Figure 1(b) shows an image of the fluorescence emission from the plasma grating, produced in atmospheric pressure air, measured using a top view camera equipped with a 10x magnification objective lens. (Note that in this measurement one of the pump laser was focused using a f=500 mm lens therefore the spot size was smaller). The image clearly shows intensity modulation of the plasma emission, and the peak intensity regions indicate the locations where the neutral gas (in this case, air) was ionized. The spacing of the intensity pattern is 3.6 μm, as expected from the 12.8° intersecting angle. A lateral lineout of the image in Fig. 1(b) taken at $z = 250$ μm is shown in Fig. 1(c). Note that the intensity modulation sits on top of a background illustrated by the red dashed line. This is a time integrated picture of the plasma emission from a combination of bremsstrahlung (free-free), recombination (free-bound) and line (bound-bound) radiation that occurs over tens of nanosecond timescale. It's interesting that the evidence of the ionization produced by the interference can still be seen in this image. To obtain Fig. 1 (d) the background was obtained by smoothing the measured intensity profile and it qualitatively represents the averaged intensity distribution of the interfering pump pulses. The ratio between the measured fluorescence intensity (blue solid line) and the estimated background (red dashed line) gives the relative intensity modulation, which is shown in Fig. 1(d). The relative intensity modulation magnitude reaches ~10% across the majority of the grating. This, however, should not be confused with the plasma density modulation discussed later.

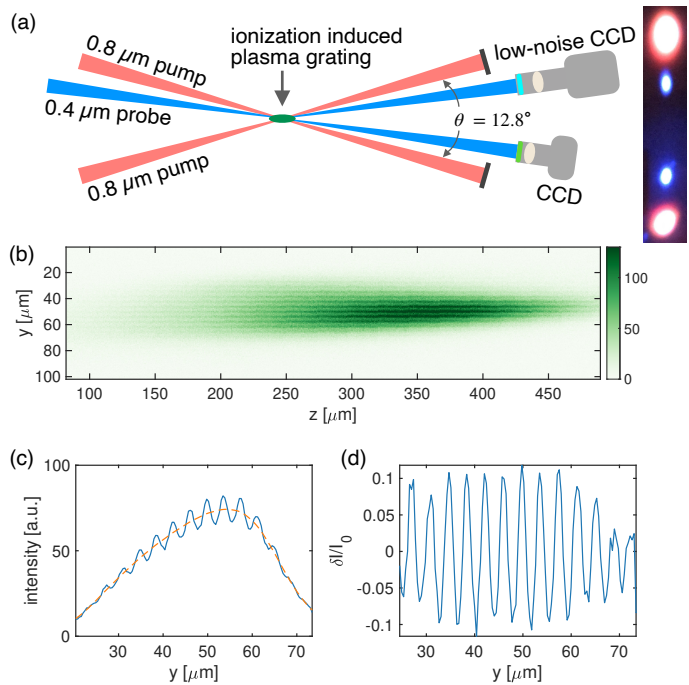

Fig. 1 Experimental setup and formation of ionization induced plasma grating. (a) shows the experimental setup. Two identical pump pulses ($\lambda_0 = 0.8$ μm, $\tau$~43 fs) intersect each other at an angle of 12.8° to form the plasma grating. A frequency doubled (0.4 μm) probe pulse is sent to the plasma grating at the Bragg angle. A photo of the spots of all four beams on a white paper is shown. The scattered and transmitted photon flux are measured simultaneously to determine the scattering efficiency. (b) is an image of the plasma fluorescence light taken using the top view camera which clearly shows the grating grooves. (c) shows a lineout of the fluorescence image where the blue line is the data and the red dashed line is the estimated background. The ratio between the blue and the red lines gives the relative intensity modulation as shown in (d).

A small portion of the pump laser was split off and frequency doubled using a 1-mm thick KDP (potassium dihydrogen phosphate) crystal to serve as the probe pulse. The probe beam was focused by a f=750 mm lens at the plasma grating with an angle of $\theta/4$ so that the Bragg diffraction condition for the first order was satisfied. When the plasma grating was present, significant amount of probe photons were scattered and collected using a low-noise scientific camera (PIXIS from Princeton Instruments). In front of the camera, a 400 nm narrow band filter (10 nm bandwidth) was used to block the scattered light with unwanted wavelengths and an f=150 mm spherical lens was used to reduce the beam size on the CCD to enhance signal to noise ratio. The majority of the probe photons were transmitted and collected by another CCD (Point Grey). In front of the camera, neutral density filters and a color filter (BG-40) were used to attenuate the transmitted probe and block the noise light, respectively, and a f=150 mm lens was used to reduce the probe beam size on the CCD chip. Both cameras were calibrated such that the number of scattered and transmitted photons can be deduced from the measured signal counts and therefore the scattering efficiency can be determined.

Once the plasma grating is formed, it can efficiently scatter the probe pulse incident at the Bragg angle. The scattering has two contributions, one from the plasma and the other from the unionized neutral gas acts as if it were a neutral refractive index grating that is $\pi$ out of phase with respect to the plasma grating. The density modulation amplitude of the plasma grating equals to the neutral gas grating and does not depend on ionization degree. Since the refractive index of neutral gas is much smaller than that of plasma, the contribution of scattering from the neutral gas grating is negligible, except at $t_0$ when the refractive index of the neutral gas grating is significantly enhanced by the instantaneous $n_2$ effect induced by the pump lasers.

The Thomson scattered power of the probe beam [37] from a periodic plasma grating is given by

$$\frac{P_s}{P_i} = \left(\frac{n_e}{n_c}\frac{kd}{4}\right)^2 \tag{1}$$

where $P_s$ and $P_i$ are the power of the scattered and incident probe, $n_e$ and $n_c$ denote electron density and the critical density for the probe beam, respectively, $k$ is the wavenumber of the probe beam and $d$ is the thickness of the plasma grating along the probe beam direction. Using this equation, the absolute plasma density $n_e$ can be deduced from the measured scattering efficiency $P_s/P_i$ for a given scattering geometry where $k$ and $d$ are known. The ionization degree $n_e/n_0$ can also be calculated if the original gas density $n_0$ is also known, for instance, by inferring from the pressure of a static fill.

**Dynamics of the plasma grating**

Once an ionization induced plasma grating (IPG) is formed, it can last for tens of ps or longer [38]. Many different mechanisms can lead to the relaxation of the plasma grating, including the ambipolar expansion of the plasma and collision-assisted free electron recombination [39]. In this section, we present measurements of dynamics of the IPG in hydrogen to illustrate the role of plasma expansion.

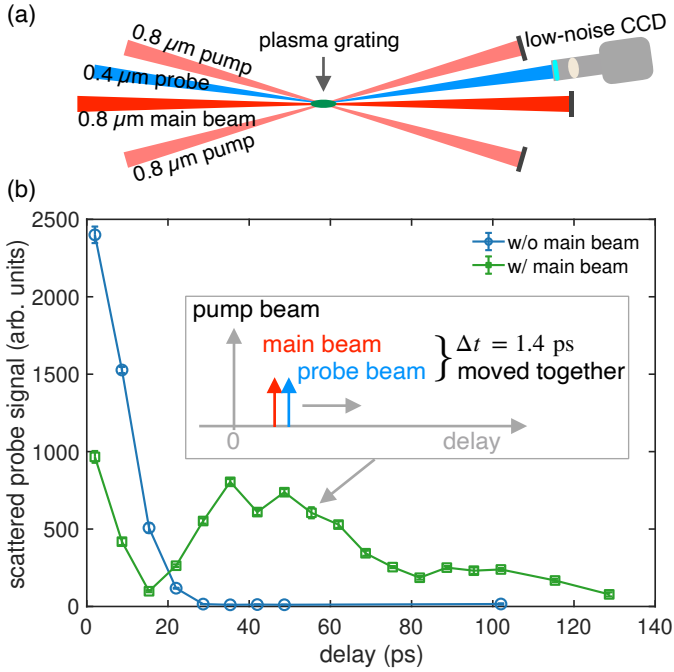

Fig. 2 Dynamics of the ionization induced plasma grating in hydrogen. (a) shows the experimental setup used for measurements of dynamics of plasma grating. In addition to the setup shown in the previous section, a 0.8 μm main beam is added to ionize the plasma grating at variable delays with respect to the pumps. (b) shows the measured evolution of the scattered probe signal without (blue circles) and with (green squares) the main beam. The data with the presence of the main beam was taken by changing the delay of the main beam and probe beam while maintaining their separation, as illustrated by the inset.

Figure 2 shows the evolution of the scattered probe signal as a function of delay with respect to the pump beams. The blue circles show the scattered probe signal when only the two pump beams were used. The density modulation magnitude is the largest when the plasma grating is just created. Thereafter, the density modulation magnitude starts decreasing and so does the scattered probe signal. The data shows an approximately exponential decay with a damping constant of 11 ps. The signal then drops to a very low level after 30 ps.

There are two possibilities for the fast damping of the ionization induced plasma grating. One is the ambipolar expansion of the plasma. Once released, the electrons start moving from the ionized to neutral region of the grating driven by thermal pressure. The charge separation then sets up an electric field which drags the heavier ions. This means that both electrons and ions from the ionized region move towards the neutral region on either side reducing the amplitude of the density modulation of the grating until it eventually vanishes. The other possibility is the collision-assisted recombination. As the free electrons recombine with the ions, the plasma grating structure also vanishes.

There is a distinguishable difference between these two scenarios. In the first case, at the time when the scattered probe signal disappears, the structure of the plasma-neutral mixture can be thought of as the superposition of a uniform density plasma and a neutral gas grating (see Fig. 3 in the next section). The unionized neutral gas structure is $\pi$ out of phase compared to the original plasma grating since the molecules can be treated as stationary on ps time scale. In the second case where recombination dominates, at the time when the scattered probe signal vanishes, the density profile of the neutral gas returns back to uniform because the previously ionized regions now have become neutral again due to recombination. The role of ambipolar expansion in the dynamics of the plasma grating can be investigated by ionizing the neutrals yet again using a third laser pulse [the "main beam" shown in Fig. 2(a)]. In the former case, when the plasma-neutral mixture is further ionized, a new plasma grating structure will form due to the ionization of the remaining neutral gas grating. On the contrary, in the latter case, no plasma grating structure is expected when the uniform neutral gas is ionized.

Experimentally we did this by sending a third 800 nm ionizing pulse ~1.5 mJ [the "main beam" pulse in Fig. 2(a)] followed by the probe pulse at variable delays with respect to the two pump beams. The peak focused intensity of the main pulse was calculated to be $\sim 7 \times 10^{14} \text{W/cm}^2$. The separation between the main beam and the probe was fixed at 1.4 ps. The measured scattered probe signal is shown by the green squares in Fig. 2(b). The first three points show a similar trend as the data without the main beam, i.e., the scattered probe signal drops with time, but the absolute magnitude of the signal is about 40% of the data without the main beam. Ideally, the signal should have dropped to zero if the main beam fully ionized the neutral gas since in that case the plasma grating is completely removed. In other words, the presence of the scattered probe signal when the main beam was present suggests that the plasma-neutral mixture was not completely ionized by the main beam due to the limited amount of energy. Nevertheless, the most important feature in the data is the emergence of the second broader peak at about 40 ps [green curve in Fig. 2 (b)]. At this delay, without the presence of the main beam, the plasma grating structure has completely disappeared as evidenced by the signal approaching zero [the blue curve in Fig. 2(b)]. However, when the main beam was present, the scattered probe signal reappeared which suggests that a second plasma grating was created. This recurrence feature indicates that the disappearance of the original plasma grating generated by the two pump beams is not dominated by recombination of the plasma, but by the expansion of the plasma.

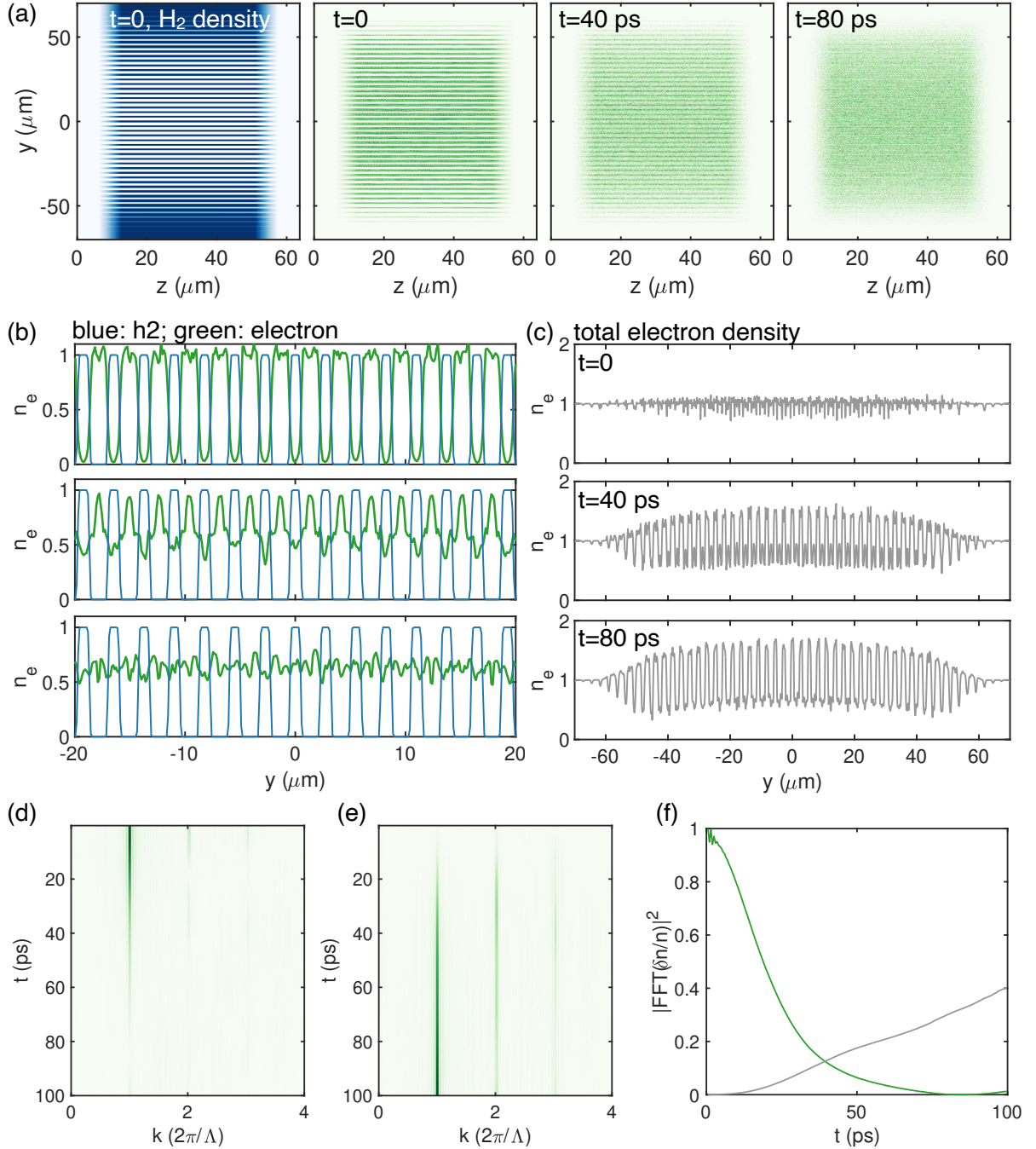

Fig. 3 PIC simulation of the dynamics of ionization induced plasma grating. (a) shows the neutral (first frame) and electron (last three frames) density at different delays. The lateral density lineouts of the neutral gas and electrons are shown in (b), represented by the blue and green lines, respectively. (c) shows the overall plasma density profiles assuming the gas-plasma mixture being fully ionized. The spatial harmonics as functions of time are shown in (d) without and (e) with the main beam, respectively. The square of the density modulation magnitude of the fundamental harmonic in both cases are shown in (f), where the green curve represents the evolution of the plasma grating without the main beam whereas the grey line shows the case with the main beam.

To further validate this physical picture suggested by the experimental data of the temporal dynamics of the ionization-grating and gain more insights, we have performed two-dimensional (2D) particle-in-cell (PIC) simulations using the Osiris framework [40] to track the evolution of the plasma grating. Ionization of neutral atoms is calculated using the

Ammosov-Delone-Krainov (ADK [41]) formulism. The simulation parameters are as follows. A 2D stationary box with length of 500 $c/\omega_0$ and width of 2000 $c/\omega_0$ was divided into 1500 by 2000 cells. Two identical laser pulses ($\lambda_0 = 0.8$ μm, $\tau_{FWHM} = 50$ fs, $w_0 = 50$ μm, $I = 1.3 \times 10^{14}$ W/cm$^2$) were injected from the left boundary of the simulation box and focused to the center of the box with an intersecting angle of 16.6°, which creates a grating with spacing of $\Lambda = 2.8$ μm. Neutral hydrogen atoms with density of $10^{17}$ cm$^{-3}$ were initialized inside the box, with four particles initialized in each cell. The hydrogen density was uniform along the transverse ($x_2$) direction and had 6.4 μm linear ramps on both side along the longitudinal ($x_1$) direction. Beyond the ramps, two 6.4 μm gaps were set to separate the hydrogen gas and the box boundaries. After being ionized, both the ions and electrons were allowed to move but the binary Coulomb collisions were not included.

In Fig. 3(a), we show snapshots of the neutral gas (the first frame) and electron density distributions taken at 0 ps, 40 ps and 80 ps (the last three frames). The lateral lineouts (integrated over 4 μm along z) of these snapshots are shown in Fig. 3(b). The three frames in Fig. 3(b) correspond to the three delays illustrated in Fig.3(a). The blue line represents the neutral density profile which is not changing and the green lines show the electron density profile that evolves with time. It is obvious that at later delays the gaps between the unionized regions have been partially filled up by the expanding electrons (and also ions) such that by 80 ps the plasma density profile becomes almost uniform. Fig. 3(c) shows what happens if, in addition to the two pump lasers, the main laser is incident upon this plasma grating at 0, 40 and 80 ps. As expected, if the main laser comes right after the pump pulses that form the grating, the fully ionized plasma will have a uniform density as shown in the top frame in Fig. 3(c). However, if the main beam ionizes the residual neutral gas later, a new plasma grating with the same period but slightly different shape is formed due to the fact that some electrons and ions have moved into the regions where no plasma existed when the original grating were created. The density modulation is seen to be up to 70%. Such a structure, when imposed onto a density downramp, can be used to periodically trigger injection of electrons into plasma wakes to form bunched electrons [42].

The evolution of the first three spatial harmonics of the plasma grating, with or without the presence of the main beam are plotted in Fig. 3(d) and (e), respectively. In both cases, the first harmonic/fundamental component dominates. In the case where the main beam is present, the higher order harmonics are more prominent. We note that Fig. 3(e) is obtained from the Fourier transform of the synthetic data in (c) instead of sending the main laser pulse in a self-consistent PIC simulation. The Thomson scattered probe signal measured in experiments is proportional to the square of the density modulation magnitude at $k = 2\pi/\Lambda$, which can be deduced from (d) and (e) and is shown in Fig. 3 (f). The green curve shows the synthetic signal of the scattering from the plasma grating without the main beam and the grey line shows the signal when the main beam is present at different delays. These curves qualitatively reproduce the observation (see Fig. 2), therefore confirm the hypothesis we made in previous paragraphs. Quantitatively, the green curve shows two times longer lifetime of the IPG and the grey curve continues grow instead of showing a similar behavior as the experimental data at large delays. These may be due to the fact that in both cases e-e or e-i and e-neutral molecules collisions as well as recombination are not included in the simulation. Both collisions and recombination would be expected to damp the plasma grating amplitude at later times. Moreover, the neutral gas molecules were considered immobile in the simulation but if their motion were included, the grey curve in Fig. 3(f) should roll off after reaching saturation.

The time evolution of the ionization grating shown in this section is based upon relative measurements of the scattered probe signal. As we have already explained, it is also possible to measure both the scattered and transmitted probe signal so as to get the scattering efficiency. Using the measured scattering efficiency, the electron density in the plasma grating can be deduced. In the following sections, we show two examples to demonstrate the usefulness of ionization induced plasma grating in measuring extremely low-density plasma. The ability to measure extremely low-density plasma may enable new possibilities in studies of strong-field ionization physics and accurate characterization of low-density (gas jet-based) plasma source essential for high-energy plasma-based accelerators.

**Absolute measurements of ionization degree of plasma grating in various gases**

Strong field ionization is the very first step of many interesting phenomena in atomic, molecular and optical physics such as above-threshold-ionization (ATI), high-order harmonic generation (HHG), nonsequential ionization (NSI) and spin-dependent ionization. In addition, optical field ionization of dense gases produces plasmas that can be an ideal platform for studying various plasma kinetic instabilities [43,44], coupling of orbital angular momentum to plasmas [45], generation of self-magnetized plasmas with spin-polarized electrons etc. We believe that ionization gratings could become a very powerful platform for studying these effects. The first step towards this is to make absolute measurements of the ionization degree of plasma grating in various atomic and molecular gases. [46].

The usual and well established method for measuring ionization degree as a function of laser intensity is by using a time-of-flight mass spectrometer [47]. In this method the relative yield of ions of a certain charge/mass ratio is measured as a function of laser intensity. Such measurements are usually performed using very low gas density, essentially in the single atom (molecule) regime [48]. While this approach is suitable for studying ionization physics of very small number of atoms (molecules), it does not allow direct ionization measurements of dense gases where interparticle spacing can be smaller than the spread of the ionizing electron wave-packet. A recent theoretical work suggests that interference of electron wave packets originating from neighboring atoms may lead to enhanced ionization for small enough interparticle distance [49]. To access this regime, the number density, or equivalently the pressure of the gas medium should be close to or above one atmosphere. We show in the following that IPG is particularly suitable for measuring the ionization degree of high density (pressure) gas medium and therefore provide new possibilities of experimentally exploring this new regime.

The measurements were performed using the setup shown sketched in Fig. 1(a). Two identical, linearly polarized pump beams, were used to create an ionization induced plasma grating in a vacuum chamber. The chamber was filled with either helium or molecular hydrogen at different pressure. A frequency doubled probe beam was focused to the plasma grating and got scattered by the free electrons in the grating. The energy of the probe beam was <1 µJ to avoid perturbing the plasma grating. The accurate (≤50 fs) synchronization of the probe beam to the pumps was established using the instantaneous $n_2$ effect of the gas. At low intensities, the two pump beams can generate a phase grating by changing the refractive index of the medium via instantaneous $n_2$ effect and such a grating also strongly diffracts the incident probe beam. This diffraction from the $n_2$ grating only happens when the probe beam was synchronized with the two pumps in helium case since the delayed $n_2$ response in this case is negligible. In molecular gases such as oxygen, nitrogen and even hydrogen, however, there exist multiple delayed $n_2$ response echoes after the pump beams have gone. The periods between echoes for nitrogen and oxygen are 8.4 ps and 11.6 ps, respectively [50], which were confirmed using our setup. Therefore, we have put the delay of the probe at ~0.5 ps for the ionization measurements to avoid the instantaneous $n_2$ effect. According to [50], the delayed

$n_2$ effects in the hydrogen case should be negligible for the actual 43 fs laser pulse used in this measurement, and indeed we did not observe significant contribution from the delayed $n_2$ response. On the other hand, the plasma grating structure should have negligible change for the 0.5 ps delay according to the grating dynamics measurements showed in the previous section.

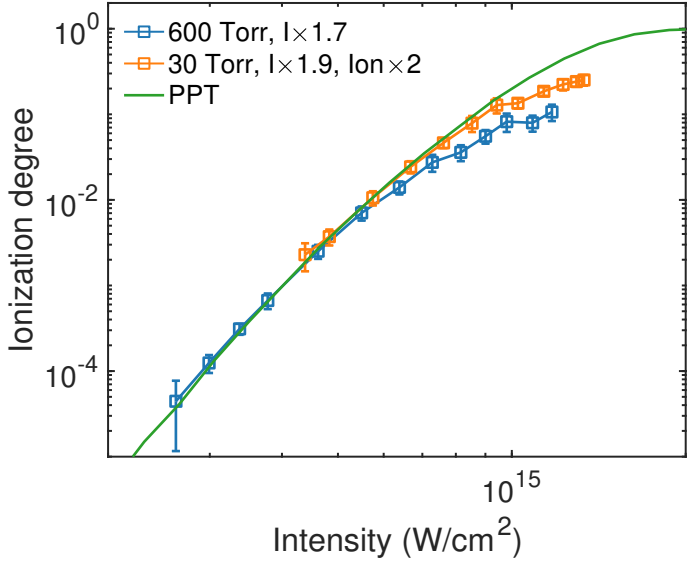

Fig. 4 Measured ionization degree of helium as a function of laser intensity. The data are shifted to match the PPT curve (see text).

The scattering efficiency was measured as a function of laser intensity and used to deduce the electron density $n_e$ using Eqn. (1). The ionization degree was then calculated as $n_e/n_0$ where $n_0$ is the number density of the neutral particles (helium atoms or hydrogen molecules) inferred using the static fill pressure, assuming each particle being singly ionized. Both the ionization degree $\delta n_e/n_0$ and the plasma grating size $d$ in Eqn. (1) depend on the laser intensity, which we write as $I_{\text{eff}} = \alpha I$, where $I_{\text{eff}}$ is the effective intensity that contributes to the formation of the plasma grating, $I$ is the intensity of each pump beam estimated using the energy, spot size and pulse duration measured in air, and $\alpha$ is a coefficient to account for factors that may reduce the intensity. For example, for two perfectly synchronized, colinear plane waves, $\alpha$ reaches its maximum value of 4; in other cases, $\alpha < 4$. In our measurements, there are several possibilities that can reduce $\alpha$. For instance, it has been reported [51] and was also observed in our experiment that both pump beams can be significantly self-scattered by the plasma grating they form. The plasma grating also enhances the spectral modulation and super-continuum generation (self and cross phase modulation and ionization-induced blue shift effects), therefore transferring energy to photons with different frequencies [52]. Both the change of the direction and the frequency of the photons reduces $\alpha$ by degrading the coherence of the two pump pulses. For example, $\alpha$ will drop by a factor of two if each pump beam loses 30% of its energy due to self-diffraction and spectral modulation. Due to the finite pulse length ($\tau \sim 43$ fs) and spot size ($w_0 \sim 45$ μm), imperfect synchronization and misalignment also causes reduction of $\alpha$. Another factor that may lead to decrease of $\alpha$ is pump depletion. The Rayleigh length of the pump pulse is $z_R \approx 8$ mm. Therefore, before the pump lasers reach the interaction point, it may have lost significant amount of energy. This was evidenced by the long (~cm) plasma filament observed in the experiment.

The width of the plasma grating also depends on laser intensity. At higher intensity, the laser ionizes a broader region which in turn increases the scattering efficiency of the plasma grating.

To take this into account, we have calculated the width of the ionization contour as a function of $I_{\text{eff}}$, and used this width as an estimation of the plasma grating size $d$ for calculating the ionization degree using Eqn. (1).

The results are summarized in Fig. 4 and 5 for helium and molecular hydrogen, respectively. To suppress self-focusing and pump depletion issues at high intensities, we used lower pressures whereas high pressure was used at low pump intensities to extend the ability of measuring low ionization degree. The blue and orange squares in Fig. 4 represent the high- and low-pressure data, respectively and the green curve is the theoretical prediction of the PPT model. To match the experimental data to the theoretical curve, we have used $\alpha = 1.7$ (high-pressure data) and 1.9 (low-pressure data) to correct the laser intensity for these datasets. The low-pressure data was stitched to the high-pressure data by aligning regions where the two datasets share the same intensity range, which required to multiply the calculated ionization degree by a factor of 2. This correction may originate from the possible change of the grating size due to shift of focus, ionization induced defocusing and misalignment between the pump and probe beams when changing the pressure.

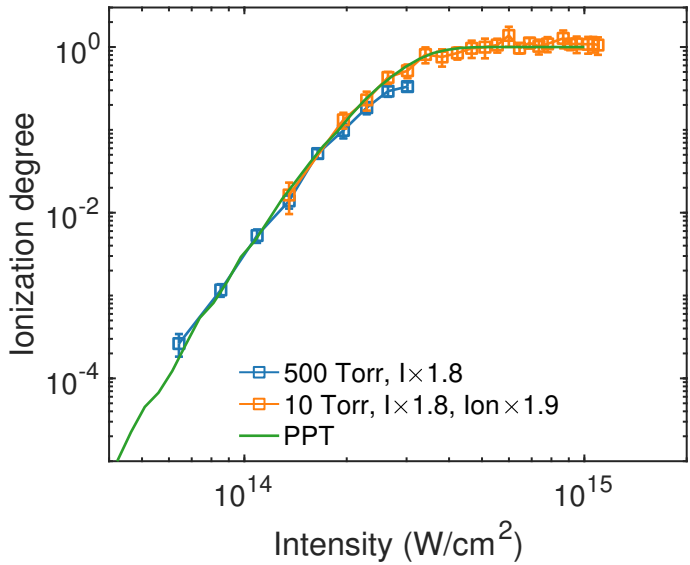

Fig. 5 Measured ionization degree of molecular hydrogen as a function of laser intensity. The green curve is the calculation using the PPT model with IP=15.43 eV and effective residual charge Z=0.95 for molecular hydrogen (see text).

The results of ionization of the simplest molecule- hydrogen using the same setup, are shown in Fig. 5. In the strong electric field of an ultrashort laser pulse there can be several different pathways for ionization of H$_2$ to eventually form 2H$^+$ + 2$e^-$. The first step is to ionize the hydrogen molecule to form $H_2^+$ ions. In the PPT calculation, we used the ionization potential of 15.43 eV for molecular hydrogen and an effective residual charge of $Z = 0.95$ (instead of $Z = 1$ for atomic hydrogen) to account for the difference between the ionization of molecular and atomic hydrogen. We note that a recent measurement using a cross-polarization dispersion interferometer also suggested that the integrated phase shift of a probe beam after propagating through the plasma showed an intensity dependence that can be best fitted by the PPT calculation using Z=0.95 [53], which coincides with the molecular PPT (MO-PPT) model [54,55]. Similar to the helium case, here we have also introduced a correction factor of $\alpha = 1.8$ for the laser intensity and for the low-pressure data the ionization degree was multiplied by a factor of two to best match the theoretical curve.

As can be seen in Figs. 4 and 5, the measured ionization degree (after correction) matches the calculation using the PPT model in the non-adiabatic regime (Keldysh parameter ~0.3-1) over a range from $10^{-4}$ to ~0.1 (helium) and 1 (hydrogen). In the helium case, the data deviates from the theory curve at the highest laser intensities, which may be due to the distortion of the pulse. The current setup did not allow us to online monitor the pump beam spot size and energy at the IP. In principle, however, an improved setup that allows a better determination of the laser intensity as well as the grating size at the IP may enable direct comparison between measurements and theory without introducing the correcting factors. Note that the PPT calculation shown in these figures only accounted for the ionization of single particle and did not include inter-particle (many-body) effects. This result suggests that the inter-particle contribution was not observed for ionization degree greater than $10^{-5}$ and the single particle ionization picture remains valid. It also suggests that the plasma density is not significantly affected by electrons lost before an electron sheath is formed around the plasma. We note that this is the first time that an ionization model for a single atom is shown to be applicable to a dense plasma to our knowledge.

The lowest plasma density measured using the plasma grating setup was ~$10^{15}$ cm$^{-3}$. This was limited by the sensitivity of the camera that measured the scattering light and the small number of photons in probe beam (~nJ) at low pump laser intensities since the probe was split from the pump and then frequency doubled. By increasing the probe intensity, it is possible to further improve the sensitivity to ~$10^{14}$ cm$^{-3}$.

**Characterization of extremely low-density gas jet**

In this section, we show another example of applications of our method, i.e., characterization of extremely low-density supersonic gas jets that are useful for plasma-based acceleration. Plasmas can serve as accelerating structures for charged particles when large-amplitude waves are excited in the wake of intense laser pulses or beam drivers. For laser wakefield accelerators, the maximum attainable energy of the trapped electrons is limited by several effects: laser diffraction, pump depletion and dephasing [2]. Pump depletion reduces the intensity of the drive laser as it continuously transfers energy to the wake and eventually to a level that no large amplitude wakes can be excited anymore. Dephasing refers to the fact that the velocity of the trapped electron bunch continuously increases and becomes larger than the group velocity of the drive laser so that the bunch eventually overruns the wake and enters the decelerating phase, which terminates the energy gain. In the 3D nonlinear regime, the pump depletion length is $k_p L_{pd} = \frac{\omega_0^2}{\omega_p^2} \omega_p \tau$ and the dephasing length is $k_p L_d = \frac{4}{3}\frac{\omega_0^2}{\omega_p^2}\sqrt{a_0}$. Here $\omega_0$ is the laser frequency and $\omega_p = \sqrt{e n_e^2/\epsilon_0 m_e}$ is the plasma frequency, $a_0$ is the normalized vector potential and $\tau$ is the duration of the laser, $k_p = c\omega_p^{-1}$ is the plasma wave number. It is obvious that by reducing the plasma density $n_e$, both the pump depletion and dephasing length can be increased to accelerate trapped bunches to higher energies. For instance, ~GeV electrons form LWFA have been observed using ~$10^{18}$ cm$^{-3}$ plasma density and 800 nm laser. By reducing the plasma density to ~$10^{17}$ cm$^{-3}$, the energy of the electrons has been increased to ~8 GeV in a recent experiment [56]. Further decreasing the plasma density to ~$10^{16}$ cm$^{-3}$ may enable the generation of ~100 GeV electrons in a single stage as suggested in [57]. These expected results assume a Ti:Sapphire laser driver. If one switches to longer wavelength drivers such as $CO_2$ lasers, the density required to reach a similar energy (10 GeV) is even lower and falls into the range of $10^{15}$ cm$^{-3}$ [58]. We note that, nowadays state-of-the-art beam-driven wakefield accelerators are also operated at low density regimes ($10^{14-17}$ cm$^{-3}$) [59–61].

While 10 cm scale gas jets have been developed that can produce such low neutral densities [11], one of the major challenges is developing methods that are capable of reliably measuring such low densities when supersonic gas gets are used to create a gas plum that is ionized by the laser pulse. In the previous section, we have shown measurements of plasma density down to ~$10^{15}$ cm$^{-3}$ by measuring the Thomson scattering efficiency from an ionization induced plasma grating. In this section, we show an example of characterizing a supersonic gas jet using the same method.

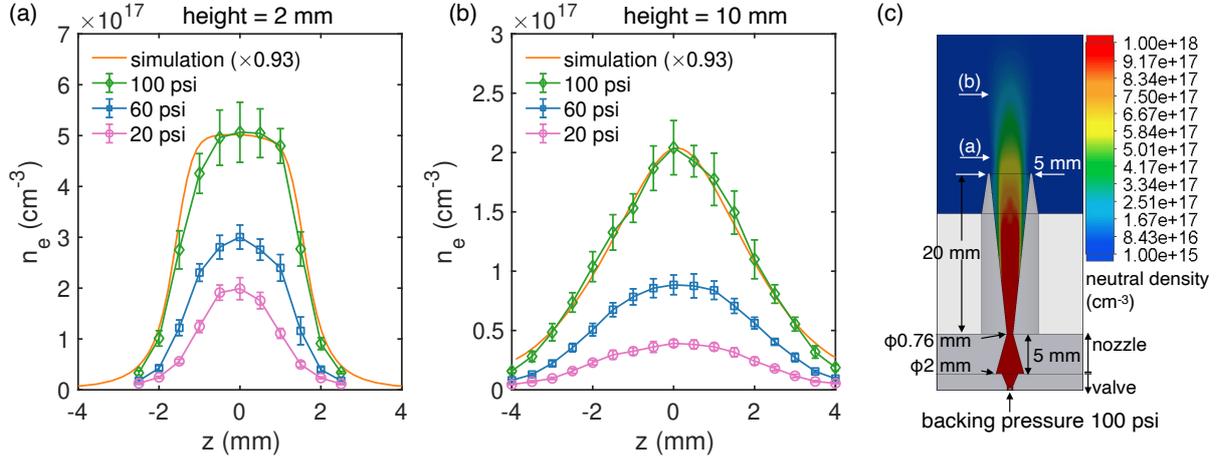

Fig. 6 Characterization of low-density supersonic gas jet. (a) and (b) show the measured plasma density at 2 mm and 10 mm above the nozzle exit, respectively. Different symbols and lines show the density profile measured using different backing pressure. Each data point is an average of 10 shots and the error bars represent the standard deviation. (c) shows a screenshot of the fluid simulation that matches the experimental conditions.

The experimental setup is the same as that in Fig. 1(a) and the results are summarized in Fig. 6. In this demonstration experiment we used a nozzle with 2-mm diameter inlet, 0.76 mm throat, and 5-mm opening to form a supersonic gas plume. The plasma density was measured at two different heights, 2 mm or 10 mm above the nozzle exit. For each height, the backing pressure of the solenoid valve was scanned from 20 to 100 psi. Under each condition, the plasma density profile was scanned by fixing the position of the plasma grating and moving the nozzle across the interaction point of the pump lasers. Figure 6(a) shows the measured plasma density profile at 2 mm. It can be seen that at relatively high backing pressure (100 psi), the density profile has a flat region across ~2 mm and a ~1 mm ramp on each side. The peak plasma density is about $5 \times 10^{17}$ cm$^{-3}$. Note that this density is already difficult to measure using conventional methods such as interferometer or wavefront sensor. When reduced backing pressure is used, the peak density drops, the width of the high-density region decreases. We note that the relative fluctuation of the density is ~10% for all different backing pressures. To further reduce the plasma density, we lowered the nozzle such that the pump lasers were focused 10 mm above the nozzle exit. The results are shown in Fig. 6(b). Compared to the 2-mm data, the noticeable changes include: the peak density drops by a factor of ~3; the width of the plasma increases and the overall shape of the density profile approaches Gaussian distribution for all backing pressures. The peak density corresponding to the lowest pressure has dropped to $3 \times 10^{16}$ cm$^{-3}$ and the density on the edge is $5 \times 10^{15}$ cm$^{-3}$. A plasma density on the order of $10^{15}$ cm$^{-3}$ is desirable for $CO_2$ laser wakefield acceleration due to the relatively long pulse duration (1-2 ps) currently available [58]. The peak density is approximately proportional to the backing pressure so that by further reducing the backing pressure down to a few psi, it is possible to lower the plasma density down to $10^{15}$ cm$^{-3}$. We also note that, these numbers represent the most accurate and the lowest density that has ever been measured using a supersonic gas jet, well below the detection limit (noise level or uncertainty) of conventional methods of about $10^{17}$ cm$^{-3}$.

To gain more insights, we have performed fluid simulations using "Solidworks". The simulation setup and the simulated neutral density map of the hydrogen flow are shown in Fig. 6(c), where the dimensions of the nozzle are labelled. The effective area of the inlet of the valve was set to 0.132 mm$^2$ (a ring belt with outer diameter of 0.76 mm specified by the manufacturer and inner diameter of 0.64 mm to account for the actual opening status of the solenoid valve), which was determined by comparing the simulated density profile with measurements using wavefront sensor at higher backing pressures (>200 psi). Then the backing pressure was reduced to 100 psi to model the low-density gas jet. The neutral hydrogen density distribution is represented by the colormap in Fig. 6(c). From this plot, we see that the density profile indeed is flatter at 2 mm height [marked by (a)] and approaches Gaussian distribution at 10 mm height [marked by (b)] due to the expansion of the gas. The neutral density profiles at these heights are shown by the orange curve in Fig. 6(a) and (b), respectively. Note that the overall shape matches the measurements remarkably. The peak density of the simulation curve was reduced by 7% to match the measurements, which is a reasonable agreement given the fact that the actual opening area of the valve was unknown and inferred from calibration using independent density measurements at high backing pressures.

The lowest density we measured using a supersonic gas jet is $5 \times 10^{15}$ cm$^{-3}$. While remarkable, we note that this is not the sensitivity limit of the diagnostic. In the previous section, we have shown that the lowest ionization degree we measured is $\sim 5 \times 10^{-5}$ for a 600 Torr static fill of helium. This converts to a plasma density of $\sim 10^{15}$ cm$^{-3}$, which represents the sensitivity of our method and is more than two orders of magnitude better than the conventional interferometer measurement. It is also worth noting that the demonstrated method measures the local plasma density within the scattering volume which has a transverse size of ~50 μm. In principle, the 3D density profile of a gas jet, not necessarily axisymmetric, can be mapped out by scanning the nozzle positions. The local measurement also avoids the artificial oscillations that often arise in density profiles deduced using Abel inversion.

**Conclusion**

Both the generation of ionization induced plasma grating using interfering lasers and their detection by Thomson scattering are well-established techniques. However, the combination of these two parts has triggered new possibilities- we have demonstrated that the frequently used model known as the PPT model for ionization of single atoms/molecules can be extended to dense plasmas in the strong field regime. Large amplitude ion gratings are shown to form by suitable arrangement of three ultrashort laser pulses. We have also shown the applicability of the collective Thomson scattering off the plasma electron gratings to measure the density profiles of extremely low-density gas jets used for plasma–based acceleration.

**Acknowledgments**

Simulations were performed on the Cori clusters at National Energy Research Scientific Computing Centre (NERSC). This work was supported by NSF Grant No. 2003354 and DOE grant DE-SC0010064.